\documentclass[12pt,preprint]{aastex}
\usepackage{natbib}

\newcommand{\gr}{$\gamma$-ray }

\newcommand{\bea}{\begin{eqnarray}}
\newcommand{\ena}{\end{eqnarray}}
\newcommand{\be}{\begin{equation}}
\newcommand{\en}{\end{equation}}
\newcommand{\bed}{\begin{displaymath}}
\newcommand{\ed}{\end{displaymath}}

\newcommand{\pd}[2]{{{\partial #1}\over {\partial #2}}}
\shorttitle{Magnetic filaments in SNR}
\shortauthors{Pohl et al.}
\begin{document}
\title{Magnetically limited X-ray filaments in young SNR}

\author{M. Pohl}
\affil{Department of Physics and Astronomy, Iowa State University, Ames, Iowa 50011-3160, USA}
\email{mkp@iastate.edu}
\author{H. Yan and A. Lazarian}
\affil{Department of Astronomy, University of Wisconsin, 5534 Sterling Hall, 
475 North Charter Street, Madison, WI 53706}

\begin{abstract}
We discuss the damping of strong magnetic turbulence downstream of the forward shock of young
supernova remnants (SNR). We find that strong magnetic fields, that have been produced by the 
streaming instability in the upstream region of the shock, or by other kinetic instabilities at
the shock, will be efficiently reduced, so the region of 
enhanced magnetic field strength would typically have a thickness of the order 
$l_d=(10^{16}-10^{17})$~cm. The non-thermal X-ray
filaments observed in young SNR are thus likely limited by the magnetic field and not by the 
energy losses of the radiating electrons. Consequently the thickness of the filaments would not 
be a measure of the magnetic field strength and claims of efficient cosmic-ray acceleration
on account of a run-away streaming instability appear premature.
\end{abstract}
%
%
\section{Introduction}
Supernova remnant shocks are prime candidates for the acceleration sites of galactic cosmic rays.
A number of young shell-type SNR have been observed to emit nonthermal X-ray emission, which is
commonly interpreted as being synchrotron emission of freshly accelerated electrons in the 10--100~TeV
energy range. These electrons, and also newly accelerated cosmic-ray nucleons, should produce
GeV--TeV scale \gr emission, which has been detected from a few objects \citep[e.g.][]{aha04}.

Likely though the inference may appear, evidence for electron acceleration in 
SNR does not imply evidence for cosmic-ray nucleon acceleration. To solve the long-standing
problem of the origin of galactic cosmic rays one is therefore interested in finding unambiguous
signatures of a large flux of high-energy cosmic-ray nucleons in SNR, which obviously requires a
good understanding of the nonthermal X-ray emission from SNR, i.e. the spatial distribution
of magnetic field and the spatial and energy distribution of high-energy electrons, so the leptonic
\gr emission can be accurately modeled. 

High-resolution X-ray observations indicate that a large fraction of the nonthermal X-ray 
emission on the rims of young SNR is concentrated in narrow filaments \citep{bamba05}.
In the literature these filaments are usually interpreted as reflecting the spatial distribution
of the high-energy electrons around their acceleration site, presumably the forward shock
\citep{yama04}. By comparing the electron energy-loss time with the downstream flow velocity one
then derives an estimate of the magnetic field strength that far exceeds the 
typical interstellar magnetic field
strength modified according to the Rankine-Hugoniot conditions \citep{vl03,bamba04}. 

A number of authors have further advanced the argument by noting that strong near-equipartition
magnetic fields can be produced by the streaming instability \citep{lb00,bl01}, thus 
apparently providing evidence for a very efficient acceleration of cosmic-ray 
nucleons at SNR forward shocks \citep{ber03a,ber03b,bv04}.

It has been pointed out before that the observed X-ray filaments may actually be magnetic filaments,
i.e. localized enhancements of the magnetic field, rather than limited by the distribution
of high-energy electrons. \citet{lp04} studied the pile-up of magnetic field at the contact 
discontinuity and concluded that it could explain the X-ray filaments, if a significant 
fraction of the electrons was accelerated close to the contact discontinuity.

In this Letter we investigate the damping of turbulent magnetic field near
SNR shocks. In contrast to earlier studies \citep{vzz88,fed92,ddk96,pz03,pz05}
that concentrated on damping in the upstream region, where 
it is in competition with the streaming instability, we only consider the downstream region
where the growth of turbulent magnetic field is negligible. We thus determine the spatial
extent of amplified turbulent magnetic field that is produced at or upstream of the forward shock.
Our analysis is independent of the amplification process, it applies to
magnetic field produced by the streaming instability in the upstream region
as well as to field generated by a kinetic plasma instability at the SNR shock
\citep{akh75}. 

Substantial progress has been made in the understanding of strong magnetic turbulence, e.g.
we have theoretical scalings that are supported by numerical simulations. Nevertheless,  
many questions remain unanswered, and in order to derive a generally valid picture,
we use various models of
the cascading and damping behaviour resulting from nonlinear wave interactions. While the models
differ in their assumptions, they all predict a significant reduction of the turbulent
magnetic energy density on a timescale shorter than that of energy losses of electrons that could 
synchrotron-radiate at a few keV X-ray energy.
Given typical plasma streaming velocities downstream of a SNR shock,
this would result in magnetic filaments of thickness $\lesssim 10^{17}$~cm.
So even if a strong magnetic field is produced at or upstream of the forward shock by any
mechanism, it is likely confined to a small volume around the shock and the observed
X-ray filaments would just reflect the structure of the magnetic fields. 

\section{Wave damping in the downstream medium of SNR shocks}
The energy density $W(k,x)$ of magnetic turbulence is related to the total
turbulent magnetic field as
\be
{{(\delta B)^2}\over {4\pi}} = \int dk\ W(k,x)
\label{1}
\en
In a plasma under
steady-state conditions $W(k,x)$ obeys the continuity equation
\be
U\,\pd{W}{x}=2\,(\Gamma_g -\Gamma_d)\,W
\label{2}
\en
where $\Gamma_g$ is the growth rate and $\Gamma_d$ is the damping rate of turbulence.
$U$ is the propagation velocity of the wave energy, i.e. the sum of the plasma
velocity and the component of the group velocity of waves along the direction of the plasma flow.
In the case under study we expect the excitation of turbulence by the
streaming instability in the upstream region at $x \le 0$, which is transported
to the downstream region ($x > 0$) and eventually damped. Alternatively one could consider
kinetic instabilities which are known to produce magnetic turbulence at the shock front
($x\simeq 0$). The instabilities
should not operate in the downstream region and hence the growth rate would 
vanish there.

Equation \ref{2} must be separately solved on either side of the shock. As boundary 
condition we may assume absence of turbulence far ahead of the shock,
$W(k,x=-\infty)=0$, whereas the MHD jump conditions applied to the upstream solution
at $x=0$ provide the boundary condition for the downstream solution at 
$x\rightarrow 0$ \citep{vs99}. 

Earlier studies found that the streaming instability during the early stages
of SNR evolution can be so strong that the amplified turbulent
magnetic field $\delta B\gtrsim 100 \ {\rm \mu G}$ far exceed the undisturbed
magnetic field $B\approx (3-10)\,{\rm \mu G}$ upstream of the SNR blast wave
\citep{lb00,bl01}.
The cosmic-ray diffusion coefficient $\kappa$ should then be close to the Bohm limit, and 
consequently both the cosmic rays with energy $E$ and the amplified magnetic field would
in the upstream region be confined to a shock precursor zone of thickness
\be
l_{\rm prec.}\approx {\kappa\over {U_s}}
\label{3}
\en
\bed
\simeq (10^{15}\ {\rm cm})\ \left({{U_s}\over {3000\,{\rm km/s}}}\right)^{-1}\, 
\left({{E}\over {\rm TeV}}\right)\, 
\left({{B}\over {100\,{\rm \mu G}}}\right)^{-1} \ed
where $U_s$ is the SNR shock velocity.

Let us in the following assume that the streaming instability has efficiently
produced strong magnetic fields in the precursor zone. The magnitude of the
amplified magnetic field upstream of the shock depends on 
the efficiency of the various possible damping mechanisms. 
In any case a fraction
of the magnetic turbulence will propagate through the shock to the downstream region,
where it will convect away from the shock. We are interested in the
spatial scale, on which the amplified magnetic energy is damped downstream of
the shock. If this length scale is small compared with the dimensions of a supernova 
remnant, then the streaming instability essentially produces a magnetic filament
at the location of the forward shock, that should be observable as a non-thermal
X-ray filament on account of enhanced synchrotron emission of very high-energy
electrons.
If, on the other hand, the damping length scale were comparable to or larger than
the typical SNR, then the entire interior of he remnant should be filled with
strong magnetic field and the observed non-thermal X-ray filaments must be limited 
by electron energy losses.

Ion-neutral collisions will not efficiently damp magnetic turbulence in the downstream
region on account of the high plasma temperature. Ordinary collisionless
damping does occur, but is very inefficient for Alfv\'en and fast-mode perturbations
that propagate along a large-scale magnetic field. Calculations of the damping rate 
of obliquely propagating fast modes for low-$\beta$ plasma \citep{gin61} and high-$\beta$
systems \citep{fk79} suggest that collisionless damping is also slow for small k or large wavelengths.
In the following we will discuss
nonlinear wave-wave interactions and cascading as potential processes
that may limit the thickness of the high magnetic field layer in the downstream region of
the SNR shock. 

The cascading of wave energy in astrophysical environments is not well
understood to date. Much of the interstellar medium, and hence the upstream medium of SNR
shocks, is thought to be turbulent. \citet{gs95} have studied the cascading of wave energy 
and found a concentration of wave energy in transverse modes, which are noted to be an inefficient 
means of cosmic-ray scattering \citep{ch00,yl02}. In addition, wave damping may occur by interactions 
with background MHD turbulence. Wave packages are distorted in
collisions with oppositely directed turbulent wave packets, resulting in a cascade of 
wave energy to smaller scaler where it is ultimately dissipated \citep{yl02}. 
Analoguous to MHD perturbations that can be decomposed into Alfv\'{e}nic, slow and fast
waves with well-defined dispersion relations, MHD perturbations that characterize turbulence 
can apparently be separated into distinct modes.

The separation into Alfv\'en and pseudo-Alfv\'en modes, which
are the incompressible limit of slow modes, is an essential element  
of the \citet{gs95} model.
Even in a compressible medium, MHD turbulence is not an inseparable mess in spite of the fact that MHD 
turbulence is a highly non-linear phenomenon \citep{LG01,CL02}. 
The actual decomposition of MHD turbulence into Alfv\'en, slow and fast 
modes was addressed in \citet{CL02,CL03}, who
used a statistical procedure of
decomposition in the Fourier space, where the basis of the Alfv\'en, slow
and fast perturbations was defined. For some particular cases, e.g. for a low $\beta$ medium, 
the procedure was benchmarked successfully and therefore argued to be reliable.

Here we list the results of three previous studies on wave damping by cascading to very small scales,
one using a very general picture of turbulence cascading, and the other two refering to the cascading of
fast-mode and Alfv\'en waves in background MHD turbulence, respectively. 

\subsection{Kolmogorov-type energy cascade}
For the case of a Kolmogorov-type nonlinearity \citet{pz03}
have considered the simplified expression
\be
\Gamma_{\rm nl}\approx {{V_A}\over {20}}\,k\,A(>k)\ ,\qquad A(>k)={{\sqrt{4\pi\,k\,W(k,x)}}\over {B_0}}
\label{4}
\en
where
$V_A$ is the Alfv\'en speed. This ansatz does not build on the disparity of the parallel and 
perpendicular scales, and therefore its applicability may be limited, but it has been used in part of 
the recent literature and thus merits our consideration.
 
In conjunction with a vanishing growth rate $\Gamma_g=0$, inserting Eq.~\ref{4}
in Eq.~\ref{2} yields
the damping length scale 
\be
l_k =20\,{{U\,\sqrt{\rho}}\over {\sqrt{W(k,0)}\,k^{3/2}}}
\label{7}
\en
with $W(k,0)$ being the energy density spectrum of turbulence that is transmitted through the 
shock to the downstream region.

Most of the magnetic energy density will reside at small $k$ or large wavelengths $\lambda$.
Inserting numbers that are typical of the downstream region of young SNR blast waves and
a highly amplified turbulent magnetic field the damping length scale is expected to be
\be
l_k\simeq (1.5\cdot 10^{16}\ {\rm cm})\,
\left({{U}\over {1000\,{\rm km/s}}}\right)\, \label{8}
\en
\bed\times\ 
\left({{n}\over {{\rm atoms/cm^3}}}\right)^{1/2}\, 
\left({{\lambda}\over {10^{15}\,{\rm cm}}}\right)\,
\left({{\delta B (\lambda)}\over {100\,{\rm \mu G}}}\right)^{-1} 
\ed
Here $\delta B$ is defined as
\be
{{\left[\delta B(\lambda)\right]^2}\over {4\pi}} =
{{\left[\delta B(k)\right]^2}\over {4\pi}} = k\, W(k,x)
\label{9}
\en
The wavelength of the turbulent magnetic field, $\lambda$, that is used in Eq.~\ref{8}
should be related to the Larmor radius $r_L$ of the high-energy particles, which are supposedly
accelerated at the shock front. Noting that
\be
r_L = (3\cdot 10^{13}\ {\rm cm})\,
\left({{E}\over {\rm TeV}}\right)\, 
\left({{B}\over {100\,{\rm \mu G}}}\right)^{-1} 
\label{10}
\en
we see that the value of $\lambda$ used in Eq.~\ref{8} corresponds to about 30 TeV 
particle energy, in the case of electrons sufficient to account for X-ray synchrotron emission
at a few keV.

For the smallest $k$ or largest wavelength $\lambda$ the magnetic fields components
used in Eqs.~\ref{8} and \ref{9} must be similar. Internal consistency should require that
the damping length scale is larger than the largest wavelength of the waves, lest the waves at 
large wavelengths would 
have very little energy density in contradiction to our assumption. In addition, an efficient backscattering 
of cosmic rays to the upstream region requires that the high magnetic field layer in the downstream region
has a thickness much larger than the scattering mean free path, which should be similar to the Larmor radius
of the particles and hence to the wavelength of the waves. This leads to a limit for the cosmic-ray energy
that is independent of the actual strength of the amplified magnetic field.
\be
l_k > \lambda\approx r_L\quad\Rightarrow\ E\lesssim 500\ {\rm TeV}
\label{11}
\en

\subsection{Cascading of fast-mode turbulence}
\citet{yl04} have studied the cascading of fast-mode turbulence, that is randomly driven by
turbulence on large scales $L$ in a compressible medium. They find a cascading timescale
\be
\Gamma_{\rm fast} \simeq \sqrt{k\over L}\,{{V_L^2}\over {V_\phi}}
\label{12}
\en
where $V_L$ is the turbulence velocity at the injection scale and $V_\phi$ is the phase velocity
of fast-mode wave and equal to the Alfv\'en and sound velocity for high- and low-$\beta$ plasma,
respectively. Typical values for the sound and Alfv\'en speed in the downstream region of the shock
of a young SNR are
\be
C_s \simeq  10^8\ {\rm cm\,s^{-1}}\label{13}
\en
\bed
V_A\simeq (2\cdot 10^7\ {\rm cm\,s^{-1}})\ \left({B\over {100\,{\rm \mu G}}}\right)
\,\left({{n_d}\over {\rm atoms\,cm^{-3}}}\right)^{-1/2}
\ed
so both characteristic velocities may be expected to be similar, corresponding to $\beta\approx 1$ in the
downstream plasma. Inserting the cascading rate (\ref{12}) as damping rate into the turbulence transport
equation (\ref{2}) we find that the turbulent magnetic field, if composed of fast-mode waves,
should exponentially decay on a length scale
\bed
l_f \simeq 
(10^{16}\ {\rm cm})\,\left({{U}\over {1000\,{\rm km/s}}}\right)\, 
\left({{V_\phi}\over {1000\,{\rm km/s}}}\right)\, 
\ed
\be
\times\ \left({{V_L}\over {1000\,{\rm km/s}}}\right)^{-2}\,
\left({{L}\over {3\,{\rm pc}}}\right)^{1/2}\, 
\left({{\lambda}\over {10^{15}\,{\rm cm}}}\right)^{1/2}\,
\label{14}
\en
where for the turbulence driving scale and velocity we have used values that are characteristic of the 
SNR blast wave itself. 

\subsection{Cascading of Alfv\'en modes}
\citet{fg04} have investigated the cascading of Alfv\'en modes by MHD turbulence that
is driven on the outer scale $L$. The damping rate is 
lowest, and the growth rate by cosmic-ray streaming is highest, for parallel-propagating Alfv\'en 
waves. All other waves with the same (parallel) wavelength should damp faster than the parallel
propagating Alfv\'en waves, so a lower limit to the damping rate of Alfv\'en waves
would be
\be
\Gamma_A\gtrsim {{V_A}\over \sqrt{\lambda\,L}}
\label{17}
\en
Again, inserting the cascading rate (\ref{17}) as damping rate into the turbulence transport
equation (\ref{2}) we find that the turbulent magnetic field, if composed of Alfv\'en waves,
should exponentially decay on a length scale
\bed
l_A \lesssim
(10^{17}\ {\rm cm})\,\left({{U}\over {1000\,{\rm km/s}}}\right)\, 
\left({{V_A}\over {500\,{\rm km/s}}}\right)^{-1}\ed
\be\times\ 
\left({{L}\over {3\,{\rm pc}}}\right)^{1/2}\, 
\left({{\lambda}\over {10^{15}\,{\rm cm}}}\right)^{1/2}
\label{18}
\en
which is very similar to the length scale obtained for the cascading of fast-mode waves in
background MHD turbulence (Eq.~\ref{14}).

\section{Discussion}
Limited and incomplete though they likely are, all models of strong turbulence cascading predict
a damping length of magnetic turbulence $l_d=(10^{16}-10^{17})$~cm that is generally small 
enough to produce a magnetic filament. The 
length scale of X-ray filaments, that are limited by electron energy losses, is larger 
than $10^{17}$~cm even for a strongly amplified magnetic field with $B\simeq 100\ {\rm \mu G}$.

To obtain a qualitative understanding of the X-ray properties of magnetic filaments we performed
a simplified calculation for a spherically symmetric SNR with a forward shock located at the radius
$r_s=10$~pc. 
The magnetic is isotropic and its spatial distribution follows
\be
B=(5\ {\rm \mu G}) + (45\ {\rm \mu G})\,\exp\left({{r-r_s}\over {0.1\ \rm pc}}\right)
\label{20}
\en
The differential electron density is at first assumed to follow
\be 
N(E)=N_0\,E^{-2}\,\exp\left(-{E\over {100\ \rm TeV}}\right)\,\Theta\left(r_s-r\right) 
\label{19}
\en
We have also studied an exponential decrease of the electron density on the lengthscale $0.5$~pc, 
i.e. $\propto \exp((r-r_s)/0.5\ {\rm pc})$, as a rough approximation to the effects of a finite 
age of the high-energy electron population.

\begin{figure}
\plotone{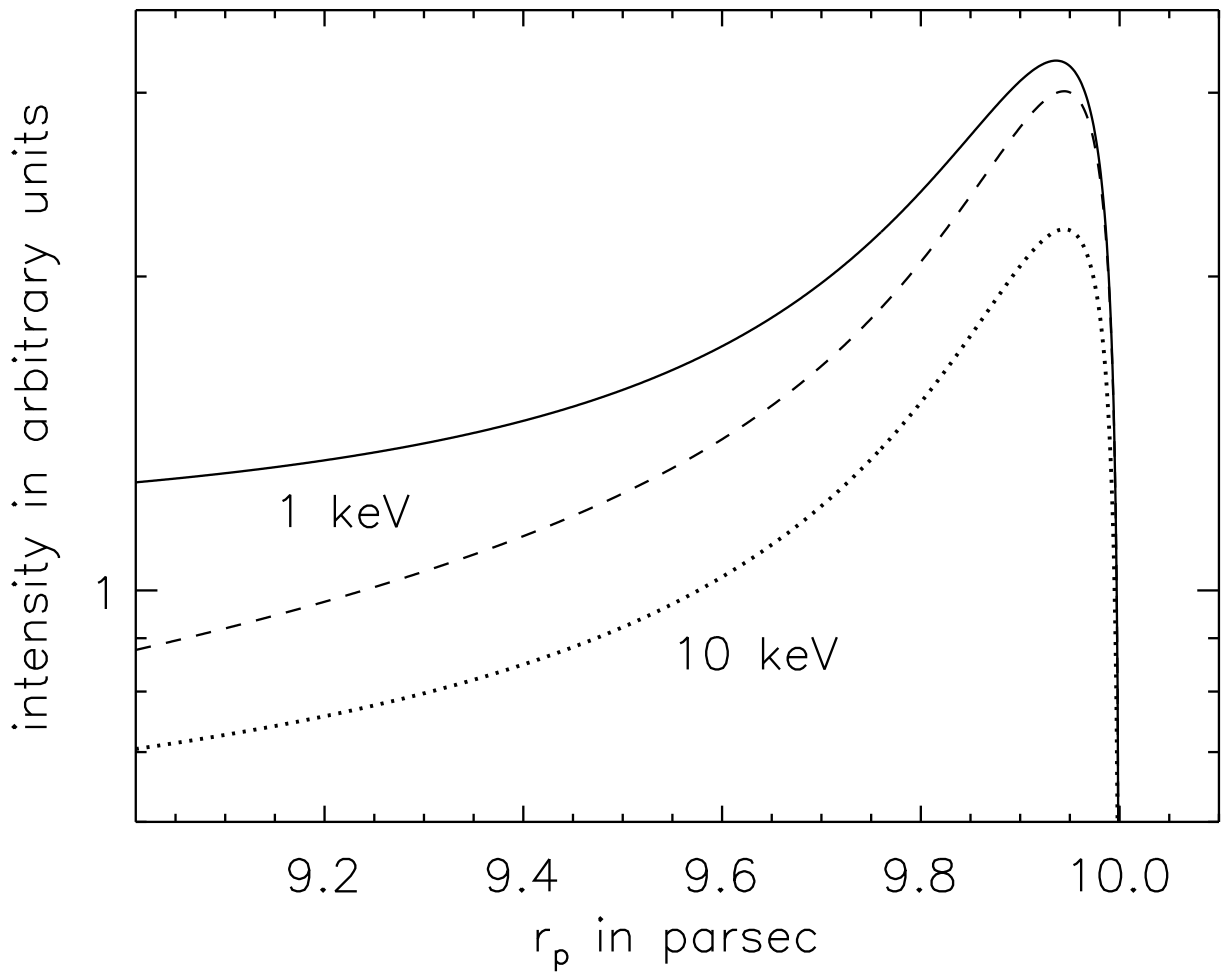}
\caption{The nonthermal X-ray intensity as a function of the projected distance from the 
SNR center. The acceleration site is assumed located at $r=10$~pc.
The solid line shows the intensity at 1 keV for a homogeneous electron density, and 
the dotted line displays the (scaled) intensity at 10 keV X-ray energy. The dashed line is derived
for the same parameters as the solid line, except that the electron density is assumed to
exponentially decay on a length scale of $0.5$~pc.}
\label{pyl-f1}
\end{figure}

The resulting X-ray intensity distribution as a function of the projected distance from the SNR center,
$r_p$,
is an integral over the synchrotron emission coefficient, $j_\nu$,
\be
I_\nu (r_p) = \int_{-\infty}^\infty dx\ j_\nu (r=\sqrt{x^2+r_p^2})
\label{21}
\en
and is shown in Fig.\ref{pyl-f1} for three cases.

To be noted from the figure is a weak frequency dependence of the downstream width of the X-ray filaments,
which is caused by the variation of the cut-off frequency of the synchrotron spectrum on account of
the spatial variation of the magnetic field strength. Spectral modeling of the radio-to-X-ray spectra of
young SNR suggests that the cut-off frequencies are generally found in the X-ray band \citep{rk97},
so we must expect some frequency dependence of the width of X-ray filaments, even if 
they are dominantly magnetic.

One should also note that any ageing or finite age of the electron population would enhance the contrast 
between the intensity of the filament and that of the plateau emission from the interior of the remnants.
For young SNR electron energy losses would be generally unimportant outside of the magnetic 
filament, though.

Our results have serious consequences for the modeling of leptonic TeV-scale \gr emission, as
that is produced anywhere, where high-energy electrons reside, and not only inside the filaments.
A careful study of the acceleration history and thus of the electron distribution
inside the remnant is required to obtain accurate estimates of the spectral energy distribution
of leptonic emission from young SNR.

\acknowledgements

Support for MP by NASA under award No. NAG5-13559 is gratefully acknowledged. 
HY is supported by the NSF grant ATM 0312282. AL acknowledge support of the Center for Magnetic Self-Organization in Laboratory and Astrophysical Plasmas.

\end{document}